\documentclass[aps,prl,reprint,nobibnotes,nofootinbib]{revtex4-1}
\usepackage{amsmath,amssymb,amsfonts,amsthm,mathtools}
\usepackage{bbm}
\usepackage{hyperref}
\usepackage{soul}
\hypersetup{colorlinks=true,linkcolor=blue,citecolor=blue,urlcolor=blue}
\usepackage{pgfplots}
\pgfplotsset{compat=1.18}
\usetikzlibrary{arrows.meta,positioning,calc}

\usepackage{tikz}
\usepackage{tikz-feynman}
\usetikzlibrary{arrows.meta,calc,decorations.pathmorphing,decorations.markings}

\begin{document}

\title{On the Limits of a Space-Time Description and the Physical Meaning of Phase Space in a Nonlocal Continuum}

\author{E.~J.~Thompson}
\thanks{Corresponding author}
\email{thom3471@mylaurier.ca}
\affiliation{Wilfrid Laurier University, Waterloo, Canada, N2L 3C5}

\date{\today}

\begin{abstract}
    In a previous paper we derived an uncertainty relation for nonlocal fields by showing that the physical localization width in nonlocal quantum field theory is broadened by the response kernel generated from the entire-function regulator. In this follow up we will reinterpret that result as just the position-sector limit of a more symmetric statement as we did not take into account the inherent nonlocality of momentum. We find and we prove under normalized, centered response kernels for both position and momentum the variance laws and derive the corresponding nonlocal phase-space uncertainty relation. The result still preserves the conclusions of the original paper all while strengthening the interpretation as nonlocal quantum field theory implies not merely a minimal measurable length, but a finite phase-space cell. We also explore experimental routes to test the theory.
\end{abstract}

\maketitle

\section{Introductory Considerations}

In local quantum theory the uncertainty principle is usually written as a relation between the spread of position and the spread of momentum and in its ordinary form this statement presupposes that both quantities may be represented, at least ideally by sharp observables associated with a point-particle phase space~\cite{Heisenberg1927,Ozawa2020,Kennard1927,Robertson1929,Schrodinger1930}. While in local quantum mechanics this may be a natural starting point, in relativistic quantum field theory, and even more so apparent in nonlocal quantum field theory this assumption becomes far too strong since the physical observables are not point-supported quantities but are defined through field operators, densities, and measurement procedures with finite spacetime support. This is the standard distributional and algebraic setting of relativistic quantum field theory where fields are operator-valued distributions and physical observables are associated with smeared fields, bounded spacetime regions, or simply averaging over spacetime, rather than with unrestricted operators at exact points~\cite{StreaterWightman1964,Haag1992,Weinberg1995,
FewsterRejzner2019,BuchholzFredenhagen2023,BohrRosenfeld1933,BohrRosenfeld1950,LandauPeierls1931,Wightman1956VEV}.

We have previously found that the overlocalized commutators for these point-particles can lead to acausality paradoxes such as Sorkin's impossible measurement, and possibly even the black hole information paradox~\cite{Sorkin1993,FewsterVerch2020,FewsterVerch2023,
PapageorgiouFraser2023,Thompson2026AsymptoticMicrocausality}. 

The physical problem underlying the uncertainty relation emerged from the early quantum theory, for example when Planck introduced the quantum of action in the black-body spectrum, Einstein applied energy quanta to the photoelectric effect, Millikan experimentally determined Planck's constant from the photoelectric relation, and Compton scattering gave direct evidence for photon momentum~\cite{Planck1901,Einstein1905,Millikan1916,Compton1923}. The terminology of the photon was subsequently introduced by Lewis in 1926~\cite{Lewis1926}. De Broglie's matter-wave hypothesis then associated momentum with wavelength, and electron diffraction provided its experimental confirmation~\cite{DeBroglie1924,DavissonGermer1927,ThomsonReid1927}.

In Ref.~\cite{Thompson2026Localization} we derived a nonlocal uncertainty relation by applying a finite response kernel induced by an entire-function regulator to the position observable, the result that we found was that the physically observed localization width obeys the variance-addition law $(\Delta X_F)^2=(\Delta X)^2+\sigma_X^2$, where \(\sigma_X\) is the nonlocal width of the nonlocal spatial response kernel, by combining this with the Heisenberg inequality gave us a lower bound on the localization width and implied a minimal length of order \(L_M\sim E_M^{-1}\) defined by the Moffat parameter. That result showed us that arbitrarily large momentum spread does not lead to arbitrarily sharp physical localization in a nonlocal field theory unlike that implied by local quantum theory.

The purpose of this paper is not to say that the previous paper was wrong, but it is rather for us to reinterpret it in a more natural and physical form, as in the earlier derivation we treated position as a nonlocal observable while keeping the momentum spread as the canonical spread of the translation generator. This is mathematically consistent and gives the correct position-sector theorem, but since the fundamental structure of nonlocal quantum field theory is a nonlocal observable algebra and if it is to be taken seriously as the theory our universe admits, then momentum measurements should also be understood nonlocally.

So we will extend the earlier work by introducing a nonlocal momentum response kernel in addition to the nonlocal position response kernel. If the measured momentum distribution is broadened by a normalized and centered kernel of variance \(\sigma_P^2\), then it obeys the corresponding variance-addition law $(\Delta P_F)^2=(\Delta P)^2+\sigma_P^2$. When put together with the position variance law and the ordinary Robertson--Heisenberg inequality this will give us a fully nonlocal phase-space uncertainty principle. We can say that the minimal nonlocal phase-space cell is enlarged from the ordinary Heisenberg cell by a nonlocal contribution, possibly leading to experimental tests.

The result we find is that nonlocal quantum field theory implies more than a minimal measurable length but rather it implies that exact point-supported phase-space data are not physically realizable in the high energy limit. So spacetime can remain a smooth Lorentz-covariant manifold, and the global Poincaré charges may remain well defined, but physical position and physical momentum are reconstructed only through nonlocal observables that are fundamental to the theory, in the same way that the Heisenberg uncertainty principle is about quantum measurements but is also fundamental to quantum theory. So this is why the earlier nonlocal uncertainty relation is the position-sector projection of a larger statement that says that the fundamental excitations of the theory are not particles at points carrying momenta at points, but nonlocal wave packets with finite phase-space resolution.

This interpretation also belongs to the longer problem of relativistic localization, originally Newton and Wigner constructed localized one-particle states, while the Foldy--Wouthuysen representation gave us the corresponding positive-energy position description for Dirac particles~\cite{NewtonWigner1949,FoldyWouthuysen1950}. The limits of these constructions were subsequently examined through axiomatic, covariant, and causal analyses of localization~\cite{Wightman1962,Fleming1965,ReehSchlieder1961, Hegerfeldt1998,Malament1996,HalvorsonClifton2002}. Unsharp localization provides an operational alternative to exact projection-valued localization~\cite{Busch1999,BuschLahtiPellonpaaYlinen2016}, while the recovery of ordinary localization only as an asymptotic or infrared notion has been considered in~\cite{Thompson2026AsymptoticLocality}.

This paper was originally meant to be a corrigendum to the previous paper where we had derived a nonlocal uncertainty relation where only the position operator was nonlocal, but as the length grew it was decided to turn it into a standalone paper.

\section{The Ordinary Wave-Mechanical Uncertainty Relation}

We can start by recalling the usual local form of the Heisenberg uncertainty principle, since the nonlocal relation derived below will be obtained by deforming the physical meaning of the variables, not by changing the algebraic proof of the uncertainty principle itself is useful to review.

We let \(\mathcal H\) be a Hilbert space, and let \(\psi\in\mathcal H\) be a normalized quantum state:
\begin{equation}
\|\psi\|^2=\langle \psi|\psi\rangle=1.
\end{equation}
We let \(X\) and \(P\) be the ordinary local position and momentum operators in one spatial dimension, we thus assume that \(X\) and \(P\) are self-adjoint operators on a common dense invariant domain \(\mathcal D\subset\mathcal H\), so that all expectation values, commutators, and variances appearing below are well defined for \(\psi\in\mathcal D\)~\cite{Weyl1927,Weyl1928,VonNeumann1932,Dirac1930}. In the usual Schrödinger representation we define the position operator \(X\) and momentum operator \(P\) acting on a wavefunction \(\psi(x)\) in the position representation of quantum mechanics:
\begin{equation}
(X\psi)(x)=x\psi(x),
\qquad
(P\psi)(x)=-i\hbar\frac{d\psi}{dx}(x),
\end{equation}
where \(x\in\mathbb R\) is the spatial coordinate and \(\hbar\) is the reduced Planck constant. These operators obey the canonical commutation relation:
\begin{equation}
[X,P]=i\hbar.
\end{equation}

For any self-adjoint operator \(A\), its expectation value in the state \(\psi\) is:
\begin{equation}
\langle A\rangle_\psi
=
\langle\psi|A\psi\rangle,
\end{equation}
and its variance is:
\begin{equation}
(\Delta_\psi A)^2
=
\left\langle\psi\left|
\left(A-\langle A\rangle_\psi\right)^2
\right|\psi\right\rangle.
\end{equation}
Now why would we calculate variance in quantum mechanics? We do this because the expectation value \(\langle A\rangle_\psi\) only gives the average outcome of measurements. Variance measures how widely individual measurements scatter around this average, so a variance of zero means the system is in an eigenstate, guaranteeing a perfectly predictable measurement. When the state is clear from context, we write simply \(\Delta A\). So for both position and momentum variance we have:
\begin{align}
(\Delta X)^2
&=
\left\langle
\left(X-\langle X\rangle\right)^2
\right\rangle,
\\
(\Delta P)^2
&=
\left\langle
\left(P-\langle P\rangle\right)^2
\right\rangle,
\end{align}
where \(\Delta X\) is the root-mean-square spread of the position distribution, and \(\Delta P\) is the root-mean-square spread of the momentum distribution.

The Robertson inequality states that for any two self-adjoint operators \(A\) and \(B\), provides a fundamental lower bound on the product of the standard deviations (\(\Delta A\) and \(\Delta B\)) of any two self-adjoint operators~\cite{Robertson1929}:
\begin{equation}
     (\Delta A)^2 (\Delta B)^2 \geq \left\vert{}\frac{1}{2i} \langle[\hat{A}, \hat{B}]\rangle\right\vert{}^2
\end{equation}
But we can just use what was originally posited by Werner Heisenberg in 1927, this states that the product of the standard deviations (or uncertainties) of two observables is bounded by the expectation value of their commutator~\cite{Heisenberg1927,Kennard1927,HilgevoordUffink2001}:
\begin{equation}
\Delta A\,\Delta B
\geq
\frac{1}{2}
\left|
\langle [A,B]\rangle
\right|,
\end{equation}
where the term \([\hat{A}, \hat{B}] = \hat{A}\hat{B} - \hat{B}\hat{A}\) explains how much the two operators interfere with each other. If they commute, the lower bound is zero, this means you can measure both simultaneously with infinite precision. One we could also use but it is not necessary is the Robertson-Schrödinger uncertainty relation, where in 1930, Erwin Schrödinger sharpened Robertson's relation by realizing that the previous bounds did not account for cases where the operators are correlated or covariant. He added a "covariance term" to the lower bound to make the inequality tighter~\cite{Schrodinger1930}:
\begin{equation}
    (\Delta A)^2 (\Delta B)^2 \geq \vert{}\langle\hat{A}\hat{B}\rangle - \langle\hat{A}\rangle\langle\hat{B}\rangle\vert{}^2 = \frac{1}{4} \vert{}\langle[\hat{A}, \hat{B}]\rangle\vert{}^2 + \vert{}\text{Cov}(A,B)\vert{}^2,
\end{equation}
with:
\begin{equation}
    \text{Cov}(A,B)=\frac{1}{2}\langle\hat{A}\hat{B}+\hat{B}
    \hat{A}\rangle-\langle\hat{A}\rangle\langle\hat{B}\rangle,
\end{equation}
where because $|\text{Cov}(A,B)|^2 \geq 0$, the Robertson-Schrödinger relation is always a stronger, more generalized condition than Robertson's inequality. Because the covariance term is squared, \(\vert{}\text{Cov}(A,B)\vert{}^2\) is always greater than or equal to zero so the Robertson-Schrödinger bound is strictly tighter whenever the two properties are correlated. It solves the "triviality problem" that states if the commutator evaluates to zero for a certain state, Robertson's bound drops to an uninformative \(\ge 0\), while the Robertson-Schrödinger relation still provides a real, non-zero boundary based on the state's covariance. It becomes an exact equality for many pure quantum states such as qubits. Basicllly if a system perfectly hits the lower bound making it an equality, it means the system has achieved the absolute minimum uncertainty physically allowed by nature. The state is said to saturate the bound. In the standard Robertson proof \(\lambda \) is to be a purely imaginary number (\(i\gamma\)). This means Robertson only hits exact equality for states where the covariance is exactly zero. Schrödinger allowed \(\lambda \) to be a complex number (\(\alpha + i\gamma\)) since it absorbs the real part (\(\alpha \)) as the covariance, the Robertson-Schrödinger relation hits an exact equality for a much wider class of pure states. A standard qubit example for a two-Level system is that for a single qubit such as a spin-$1/2$ electron or a polarized photon in any pure state, the Robertson-Schrödinger relation is always an exact equality. If you pick any two spin operators like \(\hat{S}_{x}\) and \(\hat{S}_{y}\), Robertson's formula will give you a loose lower bound that doesn't match the actual product of the variances while the Robertson-Schrödinger equation will calculate a bound that perfectly matches the actual variance product. It leaves no mathematical slack, proving that a qubit is always at the absolute geometric limit of its quantum uncertainty. Now after that tangent we can go on by applying this to \(A=X\) and \(B=P\), and using \([X,P]=i\hbar\), gives:
\begin{equation}
\Delta X\,\Delta P
\geq
\frac{1}{2}
\left|
\langle i\hbar\rangle
\right|
=
\frac{\hbar}{2}.
\end{equation}
Therefore the ordinary Heisenberg uncertainty principle is given by:
\begin{equation}
\Delta X\,\Delta P\geq \frac{\hbar}{2}.
\end{equation}

This relation is the local phase-space uncertainty principle, it refers to the ideal local variables \(X\) and \(P\), where \(X\) is interpreted as the position observable and \(P\) as the generator of spatial translations. In the local theory these variables define the usual phase-space pair \((x,p)\), and the inequality says that a quantum state cannot be made arbitrarily sharp in both variables at once~\cite{Wigner1932,Groenewold1946,Moyal1949,Hillery1984,
Folland1989,CurtrightFairlieZachos2014,MartinMartinez2022}.

A possible concern is that in canonical quantum field theory the equal-time commutation relation is not just \([X,P]=i\hbar\)~~\cite{Feynman1949,Dyson1949a,Dyson1949b}, but is:
\begin{equation}
[\phi(t,\mathbf x),\pi(t,\mathbf y)]
=
i\hbar \delta^{(3)}(\mathbf x-\mathbf y),
\end{equation}
we should ntoe that this does not contradict the present derivation as the fields \(\phi(t,\mathbf x)\) and \(\pi(t,\mathbf y)\) are operator-valued distributions, and the delta function is a contact distribution. Physical field observables must be smeared with test functions or detector profiles:
\begin{equation}
\phi(f)=\int d^3x\,f(\mathbf x)\phi(t,\mathbf x),
\\
\pi(g)=\int d^3x\,g(\mathbf x)\pi(t,\mathbf x).
\end{equation}
Then:
\begin{equation}
[\phi(f),\pi(g)]
=
i\hbar\int d^3x\,f(\mathbf x)g(\mathbf x),
\end{equation}
and the Robertson inequality gives us:
\begin{equation}
\Delta\phi(f)\Delta\pi(g)
\geq
\frac{\hbar}{2}
\left|
\int d^3x\,f(\mathbf x)g(\mathbf x)
\right|.
\end{equation}
For a normalized mode this reduces to the ordinary finite bound:
\begin{equation}
\Delta Q_f\Delta P_f\geq \frac{\hbar}{2}.
\end{equation}
So the delta function in the canonical field commutator is precisely the reason one must work with smeared observables in QFT. The present nonlocal uncertainty relation is derived at this fundamental, smeared level, where the entire-function regulator further replaces the sharp detector profile by a finite response kernel. So the proof does not assume pointlike particle observables like in the derivation above, as in the full field theory it assumes the finite Robertson bound for the smeared mode or detector observable and then proves how the nonlocal kernel enlarges the measured phase-space variances.

\section{On the Observability of Quantities in a Nonlocal Field Theory}

We now will introduce the nonlocal structure of our theory and note that the main message is that nonlocality is not added as an imperfection of the measurement, nor is it imposed by hand as a modification of the canonical commutator like in some generalized uncertainty principles~\cite{Mead1964,Garay1995,KempfManganoMann1995,
Hossenfelder2013,AmelinoCamelia2002}. It follows from the way the fundamental observables of the field theory are defined.

We let \(O(x)\) denote a local observable of the undeformed quantum field theory, where \(x\in \mathbb R^{1,3}\) is a spacetime point, and \(O(x)\) may represent a local field, a current density, an energy density, a momentum density, or a composite observable. In a local field theory, \(O(x)\) is associated with the point \(x\). In the nonlocal theory, this point-supported observable is replaced by a regulated observable, the point observable is dressed with the entire function regulator:
\begin{equation}
O_F(x)
=
F\!\left(\frac{\Box}{E_M^2}\right)O(x),
\end{equation}
where \(F\) is an entire function, \(\Box\) is the spacetime d'Alembertian, and \(E_M>0\) is the nonlocal energy scale~\cite{ReedSimon1972,Moffat1990,EvensMoffatKleppeWoodard1991,
KleppeWoodard1992,MoffatThompson2026Regulators}. We define the corresponding nonlocal length scale by $L_M:=E_M^{-1}$, in units where \(\hbar=c=1\). If units are restored, this length is of order $L_M\sim \hbar c/E_M$, the Moffat length parameter.

The function \(F\) is chosen so that the local theory is recovered in the infrared, $F(0)=1$. It is also chosen to be nonvanishing in the finite complex plane, so that the regulator does not introduce new finite poles into the propagator. In this way the nonlocal theory modifies the ultraviolet behaviour of the theory without adding unwanted extra particle states. The role of \(F\) is therefore to soften short-distance structure while preserving the ordinary low-energy limit~\cite{Feynman1948,Dyson1952,Kuzmin1989,Tomboulis1997,
BiswasMazumdarSiegel2006,BiswasGerwickKoivistoMazumdar2012,
Modesto2012,ModestoRachwal2014,TalaganisBiswasMazumdar2015,
Tomboulis2015,ModestoRachwal2017,
BuoninfanteKoshelevMazumdar2018,
BuoninfanteGiacchiniNetto2024}.

The action of \(F(\Box/E_M^2)\) may be written in terms of a kernel, we write~\cite{Schwartz1950,GelfandShilov1964,SteinWeiss1971}:
\begin{equation}
O_F(x)
=
\int d^4y\,K_F(x-y)O(y),
\end{equation}
where \(K_F(x-y)\) is the spacetime kernel associated with the entire-function regulator. This shows us explicitly what the nonlocality means, as the observable assigned to the spacetime point \(x\) is not determined only by the undeformed observable at \(x\), but by a weighted contribution from nearby spacetime points \(y\).

The nonlocality is intrinsic to spacetime as it is not the result of an inaccurate detector, but it is a structural feature of the regulated field algebra. The local observable \(O(x)\) is not itself the fundamental physical object of the nonlocal theory but the fundamental object is instead the nonlocal observable \(O_F(x)\)~\cite{PaisUhlenbeck1950,Yukawa1950a,Yukawa1950b,
RivierStueckelberg1948,UmezawaYukawaYamada1948,
Rayski1951,Blokhintsev1958,Efimov1967,Efimov1970,
AlebastrovEfimov1973,AlebastrovEfimov1974,Moffat1960}.

This has an immediate consequence for phase space, we can interpret it and say if the theory replaces point-supported spacetime observables by nonlocal observables, then it is not consistent to regard position as intrinsically nonlocal while treating momentum as completely pointlike. Position, momentum density, energy density, mass density, current density, and velocity fields all arise from the same field-theoretic observable algebra so when the algebra is regulated all such quantities must be understood as nonlocal quantities.

This does not mean that global symmetry generators are destroyed as the total four-momentum may still be well defined as the generator of spacetime translations. For example if \(T^{\mu\nu}(x)\) is the stress-energy tensor the total four-momentum is:
\begin{equation}
P^\mu
=
\int d^3x\,T^{0\mu}(t,\mathbf x),
\end{equation}
where \(T^{00}\) is the energy density and \(T^{0i}\) are the momentum densities. In the nonlocal theory we modify our equations:
\begin{equation}
T^{\mu\nu}(x)
\longrightarrow
T_F^{\mu\nu}(x)
=
F\!\left(\frac{\Box}{E_M^2}\right)T^{\mu\nu}(x).
\end{equation}
If the kernel is normalized and suitable boundary conditions hold, then the integrated charge may remain unchanged:
\begin{equation}
P_F^\mu
=
\int d^3x\,T_F^{0\mu}(t,\mathbf x)
=
P^\mu,
\end{equation}
so the global Poincaré charges may survive as exact symmetry labels, while the local densities from which they are built become nonlocal~\cite{Thompson2026SMatrix}.

This distinction will be quite important below as the nonlocal uncertainty principle is not the statement that translation invariance fails nor that the global momentum operator ceases to exist. But it is the statement that exact point supported phase-space observables are not elements of the fundamental nonlocal algebra. The local variables \(X\) and \(P\) remain useful ideal variables, but the physically fundamental quantities are the nonlocal variables \(X_F\) and \(P_F\).

\section{The Position Content of a Nonlocal Excitation}

We now will specialize the nonlocal observable algebra to the position sector, the purpose of this section is to recover the position result we found in Ref.~\cite{Thompson2026Localization}, this will be in a form that will later be extended to the momentum content of the nonlocal excitation.

We work on a fixed equal-time spatial slice, and for simplicity we first consider one spatial dimension. We will first let \(x\in\mathbb R\) denote the ordinary spatial coordinate, and then let \(p(x)\) be the ideal local position probability density associated with a normalized state, so the probability density function (PDF) for a continuous random variable is:
\begin{equation}
p(x)\geq 0,
\qquad
\int_{\mathbb R} dx\,p(x)=1.
\end{equation}
The ordinary expectation value of position is the given by:
\begin{equation}
\langle X\rangle
=
\int_{\mathbb R} dx\,x\,p(x),
\end{equation}
and the local position variance is:
\begin{equation}
(\Delta X)^2
=
\int_{\mathbb R} dx\,\bigl(x-\langle X\rangle\bigr)^2p(x),
\end{equation}
here \(X\) denotes the ideal local position variable, while \(\Delta X\) denotes the root-mean-square spread of the ideal local position distribution.

In the nonlocal theory, the physical position density is not \(p(x)\) itself as the nonlocal field algebra replaces point-supported localization by a nonlocal spatial profile. We can therefore define the nonlocal position density by:
\begin{equation}
p_F(x)
=
(\kappa_X*p)(x)
=
\int_{\mathbb R}dy\,\kappa_X(x-y)p(y),
\end{equation}
where \(\kappa_X\) is the position-space kernel induced by the nonlocal regulator on the equal-time slice. The subscript \(X\) indicates that this kernel controls the nonlocal broadening of position~\cite{BuschLahti1984,CarmeliHeinonenToigo2004}. We assume that \(\kappa_X\) satisfies the following three conditions:
\begin{equation}
\int_{\mathbb R}dx\,\kappa_X(x)=1,
\end{equation}
\begin{equation}
\int_{\mathbb R}dx\,x\,\kappa_X(x)=0,
\end{equation}
and:
\begin{equation}
\sigma_X^2
=
\int_{\mathbb R}dx\,x^2\kappa_X(x)
<\infty.
\end{equation}
The first condition says that total probability is preserved, the second condition says that the kernel is centered and therefore does not systematically shift the mean position, and the third condition defines the nonlocal position width \(\sigma_X\). This width is set by the nonlocality scale of the theory and is expected to be of order:
\begin{equation}
\sigma_X\sim L_M\sim E_M^{-1},
\end{equation}
up to a regulator-dependent numerical coefficient.

We now prove the variance law for the nonlocal position density.

\textbf{Proposition.}
Let \(p_F=\kappa_X*p\), where \(p\) is a normalized probability density and \(\kappa_X\) is normalized, centered, and has finite variance \(\sigma_X^2\). Then
\begin{equation}
(\Delta X_F)^2
=
(\Delta X)^2+\sigma_X^2.
\end{equation}

The fully worked proof can be found in Ref.~\cite{Thompson2026Localization}, but we review it here, the mean of the nonlocal position distribution is:
\begin{equation}
\langle X_F\rangle
=
\int_{\mathbb R}dx\,x\,p_F(x).
\end{equation}
Substituting the definition of \(p_F\), we obtain:
\begin{equation}
\langle X_F\rangle
=
\int_{\mathbb R}dx\,x
\int_{\mathbb R}dy\,\kappa_X(x-y)p(y).
\end{equation}
Assuming the integrals are absolutely convergent, we may exchange the order of integration:
\begin{equation}
\langle X_F\rangle
=
\int_{\mathbb R}dy\,p(y)
\int_{\mathbb R}dx\,x\,\kappa_X(x-y).
\end{equation}
Now we define:
\begin{equation}
u=x-y,
\qquad
x=u+y,
\qquad
dx=du.
\end{equation}
Then we see:
\begin{equation}
\int_{\mathbb R}dx\,x\,\kappa_X(x-y)
=
\int_{\mathbb R}du\,(u+y)\kappa_X(u).
\end{equation}
Using the normalization and centering of \(\kappa_X\), this becomes:
\begin{equation}
\int_{\mathbb R}du\,(u+y)\kappa_X(u)
=
\int_{\mathbb R}du\,u\kappa_X(u)
+
y\int_{\mathbb R}du\,\kappa_X(u)
=
0+y
=
y.
\end{equation}
Therefore:
\begin{equation}
\langle X_F\rangle
=
\int_{\mathbb R}dy\,y\,p(y)
=
\langle X\rangle,
\end{equation}
so we see that the nonlocal position kernel preserves the mean position.

We now compute the second moment because it measures the spread, scale, and variability of a distribution, it is given by:
\begin{equation}
\langle X_F^2\rangle
=
\int_{\mathbb R}dx\,x^2p_F(x).
\end{equation}
Substituting the convolution expression gives:
\begin{equation}
\langle X_F^2\rangle
=
\int_{\mathbb R}dx\,x^2
\int_{\mathbb R}dy\,\kappa_X(x-y)p(y),
\end{equation}
then again, exchanging the order of integration:
\begin{equation}
\langle X_F^2\rangle
=
\int_{\mathbb R}dy\,p(y)
\int_{\mathbb R}dx\,x^2\kappa_X(x-y).
\end{equation}
Again we set $u=x-y$, then we find:
\begin{equation}
\int_{\mathbb R}dx\,x^2\kappa_X(x-y)
=
\int_{\mathbb R}du\,(u+y)^2\kappa_X(u).
\end{equation}
Then we expand the square:
\begin{equation}
(u+y)^2=u^2+2uy+y^2,
\end{equation}
so therefore:
\begin{widetext}
\begin{equation}
\begin{aligned}
\int_{\mathbb R}du\,(u+y)^2\kappa_X(u)
=
\int_{\mathbb R}du\,u^2\kappa_X(u)
+
2y\int_{\mathbb R}du\,u\kappa_X(u)
+
y^2\int_{\mathbb R}du\,\kappa_X(u).
\end{aligned}
\end{equation}
\end{widetext}
Using the three defining properties of \(\kappa_X\), we find:
\begin{equation}
\int_{\mathbb R}du\,(u+y)^2\kappa_X(u)
=
\sigma_X^2+0+y^2
=
y^2+\sigma_X^2.
\end{equation}
So we can see the second moment as:
\begin{equation}
\langle X_F^2\rangle
=
\int_{\mathbb R}dy\,p(y)\bigl(y^2+\sigma_X^2\bigr).
\end{equation}
Since \(p\) is normalized to one, we find:
\begin{equation}
\langle X_F^2\rangle
=
\int_{\mathbb R}dy\,y^2p(y)
+
\sigma_X^2\int_{\mathbb R}dy\,p(y),
\end{equation}
so:
\begin{equation}
\langle X_F^2\rangle
=
\langle X^2\rangle+\sigma_X^2.
\end{equation}

The variance of the nonlocal position distribution is given by:
\begin{equation}
(\Delta X_F)^2
=
\langle X_F^2\rangle-\langle X_F\rangle^2.
\end{equation}
Taking:
\begin{equation}
\langle X_F\rangle=\langle X\rangle
\end{equation}
and:
\begin{equation}
\langle X_F^2\rangle=\langle X^2\rangle+\sigma_X^2,
\end{equation}
we obtain the following:
\begin{equation}
(\Delta X_F)^2
=
\langle X^2\rangle+\sigma_X^2-\langle X\rangle^2.
\end{equation}
Therefore:
\begin{equation}
(\Delta X_F)^2
=
(\Delta X)^2+\sigma_X^2.
\end{equation}
This proves the result the variace addition law.

The meaning of this equation is important, it tells us that the nonlocal position width \(\Delta X_F\) contains two contributions, the first contribution of \(\Delta X\), is the ordinary quantum spread of the underlying local position distribution. The second contribution is \(\sigma_X\), and this is the irreducible width induced by the nonlocal field algebra itself. So even if one prepares a sequence of states for where the ideal local width \(\Delta X\) becomes arbitrarily small the nonlocal width cannot fall below:
\begin{equation}
\Delta X_F\geq \sigma_X.
\end{equation}
This is the position-sector origin of the minimal localization length in nonlocal quantum field theory.

The important point for the present paper for the extension for nonlocal phase space is that this is not the full nonlocal phase-space statement. It only treats the position variable as nonlocal, if the nonlocality is a structural property of the regulated algebra then we note that the same reasoning must also be applied to momentum.

\begin{figure}[h]
\centering
\begin{tikzpicture}
\begin{axis}[
    width=0.46\textwidth,
    height=0.30\textwidth,
    axis lines=left,
    xlabel={$x$},
    ylabel={density},
    xmin=-4,xmax=4,
    ymin=0,ymax=1.05,
    samples=300,
    legend style={draw=none,fill=none,font=\small,at={(0.98,0.98)},anchor=north east},
    tick label style={font=\small},
    label style={font=\small}
]
\addplot[thick,domain=-4:4]
{1/(sqrt(2*pi)*0.45)*exp(-x^2/(2*0.45^2))};
\addlegendentry{$p(x)$}

\addplot[thick,dashed,domain=-4:4]
{1/(sqrt(2*pi)*0.65)*exp(-x^2/(2*0.65^2))};
\addlegendentry{$\kappa_X(x)$}

\addplot[thick,dashdotted,domain=-4:4]
{1/(sqrt(2*pi)*sqrt(0.45^2+0.65^2))*exp(-x^2/(2*(0.45^2+0.65^2)))};
\addlegendentry{$p_F(x)=\kappa_X*p$}

\end{axis}
\end{tikzpicture}
\caption{
The ideal local position density \(p(x)\) is not the fundamental position profile of the nonlocal theory. The regulated algebra induces a centered kernel \(\kappa_X(x)\), and the nonlocal position density is the convolution \(p_F(x)=\kappa_X*p\). Since \(\kappa_X\) is normalized and centered, the mean is preserved, while the variance is enlarged according to \((\Delta X_F)^2=(\Delta X)^2+\sigma_X^2\).
}
\label{fig:position-smearing}
\end{figure}
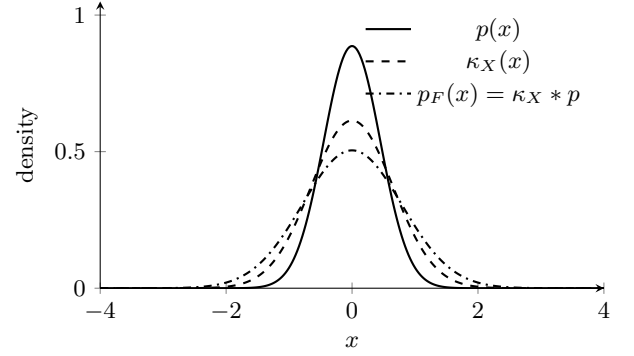

\section{The Momentum Content of a Nonlocal Excitation}

We now will apply the same nonlocal logic to momentum, this is the ingredient that I missed to put in the previous paper. In the previous section the position density was replaced by a nonlocal density because the regulated field algebra does not contain exact point supported localization observables. If the nonlocality is intrinsic to the algebra itself then the momentum content assigned to a physical excitation must also be treated as a nonlocal quantity.

We let \(p\in\mathbb R\) denote the ordinary one-dimensional momentum variable, and let \(\varrho(p)\) be the ideal local momentum probability density associated with a normalized state. So we define the two fundamental properties required for a function to be a valid probability density function, that is is positive and that the probability of finding the particle or variable at some value must be exactly \%100:
\begin{equation}
\varrho(p)\geq 0,
\qquad
\int_{\mathbb R}dp\,\varrho(p)=1.
\end{equation}
The ordinary expectation value of momentum is:
\begin{equation}
\langle P\rangle
=
\int_{\mathbb R}dp\,p\,\varrho(p),
\end{equation}
and the ordinary momentum variance is:
\begin{equation}
(\Delta P)^2
=
\int_{\mathbb R}dp\,\bigl(p-\langle P\rangle\bigr)^2\varrho(p),
\end{equation}
where \(P\) denotes the ideal canonical momentum variable, and \(\Delta P\) is the root-mean-square spread of the ideal momentum distribution.

In ordinary local quantum mechanics, \(P\) is the generator of spatial translations, and in the nonlocal theory this global role of momentum need not be abandoned. The total translation generator can remain well defined but what changes is the local and finite-resolution momentum content associated with a physical excitation. Since the underlying field observables are regulated by the nonlocal algebra then logically the momentum distribution associated with a nonlocal excitation should also be replaced by a nonlocal distribution.

We will thus define the nonlocal momentum density by~\cite{BuschLahti1984,Werner2004,CarmeliHeinonenToigo2004,
BuschLahtiWerner2014}:
\begin{equation}
\varrho_F(p)
=
(\kappa_P*\varrho)(p)
=
\int_{\mathbb R}dq\,\kappa_P(p-q)\varrho(q),
\end{equation}
where \(\kappa_P\) is the momentum-space kernel associated with the nonlocal broadening of momentum. The variable \(q\in\mathbb R\) is an integration variable representing the ideal local momentum, while \(p\) is the momentum variable of the nonlocal distribution. The subscript \(P\) on \(\kappa_P\) tell us that this kernel controls the nonlocal resolution of momentum~\cite{Werner2004}.

We assume that \(\kappa_P\) is normalized, centered, and has finite variance:
\begin{equation}
\int_{\mathbb R}dp\,\kappa_P(p)=1,
\end{equation}
\begin{equation}
\int_{\mathbb R}dp\,p\,\kappa_P(p)=0,
\end{equation}
and:
\begin{equation}
\sigma_P^2
=
\int_{\mathbb R}dp\,p^2\kappa_P(p)
<\infty.
\end{equation}
The first condition says that total probability in momentum space is preserved, the second condition says that the nonlocal momentum kernel does not systematically shift the mean momentum, and the third condition defines the nonlocal momentum width \(\sigma_P\). The parameter \(\sigma_P\) has the dimensions of momentum. These are the same conditions as in the previous section, just for the momentum observable. So since the nonlocal length scale is \(L_M\sim E_M^{-1}\), the natural momentum scale associated with the nonlocal theory is of order $E_M\sim L_M^{-1}$, in units where \(\hbar=c=1\). So \(\sigma_P\) is expected to be controlled by the nonlocal scale \(E_M\), up to a regulator-dependent numerical coefficient.

We now prove the corresponding variance law in the same fashion as in Ref.~\cite{Thompson2026Localization}.

\textbf{Proposition.}
Let \(\varrho_F=\kappa_P*\varrho\), where \(\varrho\) is a normalized momentum probability density and \(\kappa_P\) is normalized, centered, and has finite variance \(\sigma_P^2\). Then a variance addition law holds:
\begin{equation}
(\Delta P_F)^2
=
(\Delta P)^2+\sigma_P^2.
\end{equation}

The proof starts with taking the mean of the nonlocal momentum distribution as:
\begin{equation}
\langle P_F\rangle
=
\int_{\mathbb R}dp\,p\,\varrho_F(p).
\end{equation}
Substituting the convolution definition gives:
\begin{equation}
\langle P_F\rangle
=
\int_{\mathbb R}dp\,p
\int_{\mathbb R}dq\,\kappa_P(p-q)\varrho(q).
\end{equation}
Assuming the relevant integrals are absolutely convergent, we exchange the order of integration as before:
\begin{equation}
\langle P_F\rangle
=
\int_{\mathbb R}dq\,\varrho(q)
\int_{\mathbb R}dp\,p\,\kappa_P(p-q).
\end{equation}
Now we take $u=p-q, \quad
p=u+q,\quad dp=du$, then we find:
\begin{equation}
\int_{\mathbb R}dp\,p\,\kappa_P(p-q)
=
\int_{\mathbb R}du\,(u+q)\kappa_P(u).
\end{equation}
Using the normalization and centering conditions on \(\kappa_P\), we get:
\begin{equation}
\int_{\mathbb R}du\,(u+q)\kappa_P(u)
=
\int_{\mathbb R}du\,u\kappa_P(u)
+
q\int_{\mathbb R}du\,\kappa_P(u)
=
0+q
=
q.
\end{equation}
Thus:
\begin{equation}
\langle P_F\rangle
=
\int_{\mathbb R}dq\,q\,\varrho(q)
=
\langle P\rangle,
\end{equation}
so the nonlocal momentum kernel preserves the mean momentum.

We now compute the second moment as before:
\begin{equation}
\langle P_F^2\rangle
=
\int_{\mathbb R}dp\,p^2\varrho_F(p),
\end{equation}
using the convolution definition:
\begin{equation}
\langle P_F^2\rangle
=
\int_{\mathbb R}dp\,p^2
\int_{\mathbb R}dq\,\kappa_P(p-q)\varrho(q).
\end{equation}
Exchanging the order of integration which is valid under the aforementioned condition, gives:
\begin{equation}
\langle P_F^2\rangle
=
\int_{\mathbb R}dq\,\varrho(q)
\int_{\mathbb R}dp\,p^2\kappa_P(p-q).
\end{equation}
Again set we let $u=p-q, \quad p=u+q$, and then we find:
\begin{equation}
\int_{\mathbb R}dp\,p^2\kappa_P(p-q)
=
\int_{\mathbb R}du\,(u+q)^2\kappa_P(u).
\end{equation}
Expanding:
\begin{equation}
(u+q)^2=u^2+2uq+q^2.
\end{equation}
Thus:
\begin{widetext}
\begin{equation}
\begin{aligned}
\int_{\mathbb R}du\,(u+q)^2\kappa_P(u)
=
\int_{\mathbb R}du\,u^2\kappa_P(u)
+
2q\int_{\mathbb R}du\,u\kappa_P(u)
+
q^2\int_{\mathbb R}du\,\kappa_P(u).
\end{aligned}
\end{equation}
\end{widetext}
Using the previous assumptions on \(\kappa_P\), this becomes:
\begin{equation}
\int_{\mathbb R}du\,(u+q)^2\kappa_P(u)
=
\sigma_P^2+0+q^2
=
q^2+\sigma_P^2.
\end{equation}
Therefore:
\begin{equation}
\langle P_F^2\rangle
=
\int_{\mathbb R}dq\,\varrho(q)\bigl(q^2+\sigma_P^2\bigr).
\end{equation}
Since \(\varrho\) is normalized:
\begin{equation}
\langle P_F^2\rangle
=
\int_{\mathbb R}dq\,q^2\varrho(q)
+
\sigma_P^2\int_{\mathbb R}dq\,\varrho(q),
\end{equation}
so:
\begin{equation}
\langle P_F^2\rangle
=
\langle P^2\rangle+\sigma_P^2.
\end{equation}

The variance of the nonlocal momentum distribution is:
\begin{equation}
(\Delta P_F)^2
=
\langle P_F^2\rangle-\langle P_F\rangle^2.
\end{equation}
By using $\langle P_F\rangle=\langle P\rangle,$ and:
\begin{equation}
\langle P_F^2\rangle=\langle P^2\rangle+\sigma_P^2,
\end{equation}
we find:
\begin{equation}
(\Delta P_F)^2
=
\langle P^2\rangle+\sigma_P^2-\langle P\rangle^2.
\end{equation}
Therefore:
\begin{equation}
(\Delta P_F)^2
=
(\Delta P)^2+\sigma_P^2.
\end{equation}
This proves the proposition for momentum in the same sense for position.

This result should not be interpreted as saying that the global momentum generator disappears, nor that the theory necessarily contains a maximum momentum but that the canonical momentum \(P\) may still exist as the generator of translations, and the total four-momentum may remain a good symmetry charge. The statement made is that the momentum content of a physical excitation, when described by the nonlocal observable algebra, is not represented by an exact point in momentum space.

So the nonlocal theory contains two distinct momentum notions, where the first is the global translation generator, which encodes the symmetry of spacetime translations. The second is the nonlocal momentum distribution associated with a physical excitation of the regulated field algebra. The variables entering the nonlocal phase-space relation are therefore not the ideal local pair \((X,P)\), but the nonlocal pair \((X_F,P_F)\).

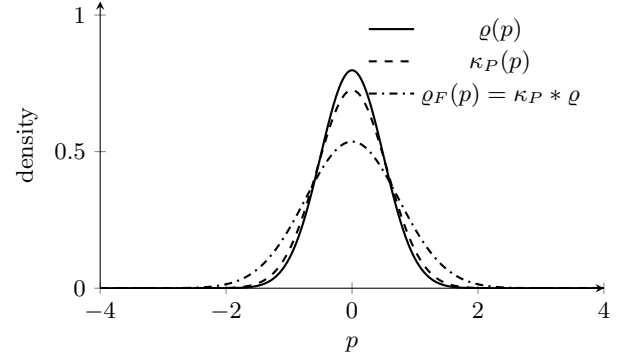
\begin{figure}[h]
\centering
\begin{tikzpicture}
\begin{axis}[
    width=0.46\textwidth,
    height=0.30\textwidth,
    axis lines=left,
    xlabel={$p$},
    ylabel={density},
    xmin=-4,xmax=4,
    ymin=0,ymax=1.05,
    samples=300,
    legend style={draw=none,fill=none,font=\small,at={(0.98,0.98)},anchor=north east},
    tick label style={font=\small},
    label style={font=\small}
]
\addplot[thick,domain=-4:4]
{1/(sqrt(2*pi)*0.50)*exp(-x^2/(2*0.50^2))};
\addlegendentry{$\varrho(p)$}

\addplot[thick,dashed,domain=-4:4]
{1/(sqrt(2*pi)*0.55)*exp(-x^2/(2*0.55^2))};
\addlegendentry{$\kappa_P(p)$}

\addplot[thick,dashdotted,domain=-4:4]
{1/(sqrt(2*pi)*sqrt(0.50^2+0.55^2))*exp(-x^2/(2*(0.50^2+0.55^2)))};
\addlegendentry{$\varrho_F(p)=\kappa_P*\varrho$}

\end{axis}
\end{tikzpicture}
\caption{
The ideal canonical momentum distribution \(\varrho(p)\) is replaced by the nonlocal distribution \(\varrho_F(p)=\kappa_P*\varrho\). The centered kernel \(\kappa_P\) preserves the mean momentum but adds its variance to the momentum width, giving \((\Delta P_F)^2=(\Delta P)^2+\sigma_P^2\). This does not remove the global translation generator; it says that the momentum content of a finite excitation is nonlocal in the regulated field algebra.
}
\label{fig:momentum-smearing}
\end{figure}

\section{The Phase-Space Limitation in the Nonlocal Theory}

We can now combine our nonlocal position and momentum variance laws with the ordinary Heisenberg inequality, so we first let \(X\) and \(P\) denote the ideal local position and momentum variables respectively. Their variances in a normalized state are given by:
\begin{equation}
(\Delta X)^2
\qquad\text{and}\qquad
(\Delta P)^2.
\end{equation}
These ideal variances obey the ordinary Heisenberg relation. The variables \(X\) and \(P\) are not taken to be the physical variables of the nonlocal theory. These are the local variables that appear before the nonlocal algebra is imposed.

The nonlocal position and momentum variables are denoted by $X_F$ and $P_F$. Here the subscript \(F\) as before indicates that these quantities are defined by the regulated field algebra generated by the entire-function operator \(F(\Box/E_M^2)\). The corresponding variances are given by:
\begin{equation}
(\Delta X_F)^2
\qquad\text{and}\qquad
(\Delta P_F)^2.
\end{equation}
From the two variance-addition laws we proved above, we have:
\begin{equation}
(\Delta X_F)^2
=
(\Delta X)^2+\sigma_X^2,
\end{equation}
and:
\begin{equation}
(\Delta P_F)^2
=
(\Delta P)^2+\sigma_P^2,
\end{equation}
where \(\sigma_X\) is the nonlocal width of the position kernel \(\kappa_X\), while \(\sigma_P\) is the nonlocal width of the momentum kernel \(\kappa_P\). The parameter \(\sigma_X\) has dimensions of length, and \(\sigma_P\) has dimensions of momentum.

Solving the two variance laws for the ideal local variances gives:
\begin{equation}
(\Delta X)^2
=
(\Delta X_F)^2-\sigma_X^2,
\end{equation}
and:
\begin{equation}
(\Delta P)^2
=
(\Delta P_F)^2-\sigma_P^2.
\end{equation}
Since ordinary variances are nonnegative, the nonlocal variances must satisfy the fact they are positive:
\begin{equation}
\Delta X_F\geq \sigma_X,
\qquad
\Delta P_F\geq \sigma_P,
\end{equation}
the equalities would correspond to vanishing ideal local spread in the corresponding variable.

We then substitute the expressions for \(\Delta X\) and \(\Delta P\) into the ordinary Heisenberg inequality gives us:
\begin{equation}
\sqrt{(\Delta X_F)^2-\sigma_X^2}\,
\sqrt{(\Delta P_F)^2-\sigma_P^2}
\geq
\frac{\hbar}{2}.
\end{equation}
Squaring both sides gives the nonlocal phase-space uncertainty relation:
\begin{equation}
\left[(\Delta X_F)^2-\sigma_X^2\right]
\left[(\Delta P_F)^2-\sigma_P^2\right]
\geq
\frac{\hbar^2}{4},
\end{equation}
this is the symmetric form of the uncertainty principle in the nonlocal theory that was not derived fully in the previous paper.

If \(\Delta P_F>\sigma_P\), we may solve for the nonlocal position width:
\begin{equation}
\Delta X_F
\geq
\sqrt{
\sigma_X^2
+
\frac{\hbar^2}{
4\left[(\Delta P_F)^2-\sigma_P^2\right]
}
}.
\end{equation}
Similarly, if \(\Delta X_F>\sigma_X\), we may solve for the nonlocal momentum width:
\begin{equation}
\Delta P_F
\geq
\sqrt{
\sigma_P^2
+
\frac{\hbar^2}{
4\left[(\Delta X_F)^2-\sigma_X^2\right]
}
}.
\end{equation}

These formulas show us explicitly how the local Heisenberg principle is embedded inside the nonlocal theory, where the ordinary uncertainty relation is not discarded. Instead we note that it applies to the underlying ideal variables \(X\) and \(P\), while the nonlocal variables \(X_F\) and \(P_F\) contain additional nonlocal widths fixed by the regulated observable algebra.

We can note some simple but important limits. The first is if the momentum nonlocality is neglected by taking:
\begin{equation}
\sigma_P\to 0,
\end{equation}
then the phase-space relation becomes:
\begin{equation}
\left[(\Delta X_F)^2-\sigma_X^2\right](\Delta P)^2
\geq
\frac{\hbar^2}{4}.
\end{equation}
Solving for \(\Delta X_F\), we find:
\begin{equation}
\Delta X_F
\geq
\sqrt{
\sigma_X^2
+
\frac{\hbar^2}{4(\Delta P)^2}
}.
\end{equation}
This is precisely the one-sided position-sector relation derived in the previous paper. So the previous result is not invalidated as it is recovered as the special case in which the momentum spread is treated as the canonical local spread.

Second, in the local limit where:
\begin{equation}
\sigma_X\to 0,
\qquad
\sigma_P\to 0,
\end{equation}
the nonlocal variables reduce to the ordinary local variables:
\begin{equation}
\Delta X_F\to \Delta X,
\qquad
\Delta P_F\to \Delta P,
\end{equation}
so the nonlocal relation then reduces to the standard Heisenberg uncertainty principle.

The ultraviolet meaning of the result is different from the local theory, it is subtle but important. In the local theory we can try to make \(\Delta X\) small by making \(\Delta P\) large, but in the nonlocal theory by increasing the momentum spread we cannot make the position spread arbitrarily small. We can see this from:
\begin{equation}
\Delta X_F
\geq
\sqrt{
\sigma_X^2
+
\frac{\hbar^2}{
4\left[(\Delta P_F)^2-\sigma_P^2\right]
}
},
\end{equation}
we see that:
\begin{equation}
\Delta P_F\to \infty
\qquad\Longrightarrow\qquad
\Delta X_F\to \sigma_X.
\end{equation}
So \(\sigma_X\) is a genuine fundamental lower bound on position resolution.

This result says more than the ordinary uncertainty principle as it tells us that exact point-supported phase-space data are not available in the fundamental nonlocal theory. The local pair \((X,P)\) is replaced by the nonlocal pair $(X_F,P_F)$, and the physical phase-space spread is bounded not only by \(\hbar\), but also by the nonlocal widths \(\sigma_X\) and \(\sigma_P\).

\section{The Finite Cell as a Condition for Physical Description}

We now derive the minimal nonlocal phase-space cell, this is natural to us as now we have both position and momentum as nonlocal. The previous section gave the symmetric uncertainty relation and this relation shows how the local Heisenberg inequality is embedded inside the nonlocal theory. In this section we want to ask a sharper question, that being what is the smallest possible value of the product $\Delta X_F\Delta P_F$? This product measures the total nonlocal phase-space spread of a state when both position and momentum are treated as nonlocal variables.

Recall the variance-addition laws:
\begin{equation}
\notag (\Delta X_F)^2=(\Delta X)^2+\sigma_X^2,
\end{equation}
and:
\begin{equation}
\notag (\Delta P_F)^2=(\Delta P)^2+\sigma_P^2,
\end{equation}
where \(\Delta X\) and \(\Delta P\) are again the ideal local position and momentum spreads, while \(\Delta X_F\) and \(\Delta P_F\) are the nonlocal spreads. The constants \(\sigma_X\) and \(\sigma_P\) are the widths of the nonlocal position and momentum kernels respectively. So \(\sigma_X\) has dimensions of length, \(\sigma_P\) has dimensions of momentum, and the product $\sigma_X\sigma_P$ has dimensions of action like \(\hbar\).

We assume that the underlying ideal variables satisfy the ordinary Heisenberg inequality. To find the smallest possible value of \(\Delta X_F\Delta P_F\) it is sufficient to minimize over states saturating the ordinary local inequality. This is because any state with:
\begin{equation}
\notag \Delta X\Delta P>\frac{\hbar}{2}
\end{equation}
has a larger underlying local phase-space spread and therefore cannot give a smaller nonlocal phase-space product. We can then write:
\begin{equation}
\Delta X=\frac{\hbar}{2\Delta P},
\end{equation}
where \(\Delta P>0\). Substituting this into the nonlocal product gives:
\begin{equation}
\notag \Delta X_F\Delta P_F
=
\sqrt{(\Delta X)^2+\sigma_X^2}\,
\sqrt{(\Delta P)^2+\sigma_P^2}.
\end{equation}
And then we see that:
\begin{equation}
\left(\Delta X_F\Delta P_F\right)^2
=
\left[
\frac{\hbar^2}{4(\Delta P)^2}
+
\sigma_X^2
\right]
\left[
(\Delta P)^2+\sigma_P^2
\right].
\end{equation}
For notational clarity, we can define:
\begin{equation}
u:=(\Delta P)^2.
\end{equation}
Since \(\Delta P>0\), we have:
\begin{equation}
u>0.
\end{equation}
Then we see:
\begin{equation}
\left(\Delta X_F\Delta P_F\right)^2
=
\left(
\frac{\hbar^2}{4u}
+
\sigma_X^2
\right)
\left(
u+\sigma_P^2
\right).
\end{equation}
Expanding this expression gives us:
\begin{equation}
\left(\Delta X_F\Delta P_F\right)^2
=
\frac{\hbar^2}{4}
+
\frac{\hbar^2\sigma_P^2}{4u}
+
\sigma_X^2u
+
\sigma_X^2\sigma_P^2.
\end{equation}
Only the middle two terms depend on \(u\). So we can therefore we minimize the function:
\begin{equation}
f(u)
=
\sigma_X^2u
+
\frac{\hbar^2\sigma_P^2}{4u},
\qquad
u>0.
\end{equation}
Taking the derivative gives us:
\begin{equation}
\frac{df}{du}
=
\sigma_X^2
-
\frac{\hbar^2\sigma_P^2}{4u^2}.
\end{equation}
The minimum occurs when:
\begin{equation}
\frac{df}{du}=0,
\end{equation}
so:
\begin{equation}
\sigma_X^2
=
\frac{\hbar^2\sigma_P^2}{4u^2}.
\end{equation}
Solving for \(u\), we find:
\begin{equation}
u
=
\frac{\hbar\sigma_P}{2\sigma_X}.
\end{equation}
Since \(u=(\Delta P)^2\), the minimizing ideal momentum spread is:
\begin{equation}
(\Delta P)^2_{\rm min}
=
\frac{\hbar\sigma_P}{2\sigma_X}.
\end{equation}
Equivalently:
\begin{equation}
\Delta P_{\rm min}
=
\sqrt{\frac{\hbar\sigma_P}{2\sigma_X}}.
\end{equation}
The corresponding ideal position spread is:
\begin{equation}
\Delta X_{\rm min}
=
\frac{\hbar}{2\Delta P_{\rm min}}
=
\sqrt{\frac{\hbar\sigma_X}{2\sigma_P}}.
\end{equation}

We can now evaluate the nonlocal product at the minimum, at the minimizing value of \(u\):
\begin{equation}
\sigma_X^2u
=
\sigma_X^2
\left(
\frac{\hbar\sigma_P}{2\sigma_X}
\right)
=
\frac{\hbar\sigma_X\sigma_P}{2},
\end{equation}
and:
\begin{equation}
\frac{\hbar^2\sigma_P^2}{4u}
=
\frac{\hbar^2\sigma_P^2}{4}
\left(
\frac{2\sigma_X}{\hbar\sigma_P}
\right)
=
\frac{\hbar\sigma_X\sigma_P}{2}.
\end{equation}
Therefore:
\begin{equation}
\left(\Delta X_F\Delta P_F\right)^2_{\rm min}
=
\frac{\hbar^2}{4}
+
\frac{\hbar\sigma_X\sigma_P}{2}
+
\frac{\hbar\sigma_X\sigma_P}{2}
+
\sigma_X^2\sigma_P^2.
\end{equation}
Combining terms gives:
\begin{equation}
\left(\Delta X_F\Delta P_F\right)^2_{\rm min}
=
\frac{\hbar^2}{4}
+
\hbar\sigma_X\sigma_P
+
\sigma_X^2\sigma_P^2.
\end{equation}
The right-hand side is a perfect square:
\begin{equation}
\frac{\hbar^2}{4}
+
\hbar\sigma_X\sigma_P
+
\sigma_X^2\sigma_P^2
=
\left(
\frac{\hbar}{2}
+
\sigma_X\sigma_P
\right)^2.
\end{equation}
Taking the positive square root, and noting thaat since uncertainties are nonnegative, this gives us:
\begin{equation}
\Delta X_F\Delta P_F
\geq
\frac{\hbar}{2}
+
\sigma_X\sigma_P,
\end{equation}
this is the nonlocal phase-space cell.

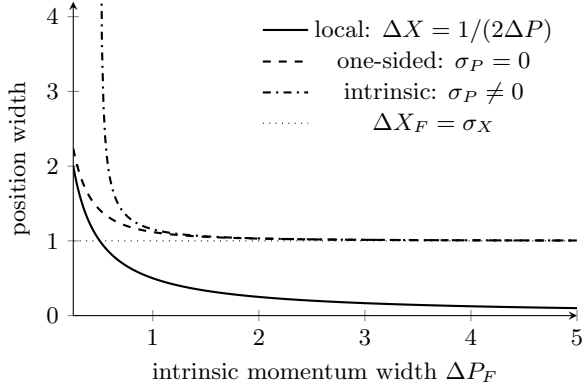
\begin{figure}[h]
\centering
\begin{tikzpicture}
\begin{axis}[
    width=0.46\textwidth,
    height=0.32\textwidth,
    axis lines=left,
    xlabel={intrinsic momentum width \(\Delta P_F\)},
    ylabel={position width},
    xmin=0.25,xmax=5,
    ymin=0,ymax=4.2,
    samples=300,
    legend style={draw=none,fill=none,font=\small,at={(0.98,0.98)},anchor=north east},
    tick label style={font=\small},
    label style={font=\small}
]
\addplot[thick,domain=0.25:5]
{1/(2*x)};
\addlegendentry{local: \(\Delta X=1/(2\Delta P)\)}

\addplot[thick,dashed,domain=0.25:5]
{sqrt(1 + 1/(4*x^2))};
\addlegendentry{one-sided: \(\sigma_P=0\)}

\addplot[thick,dashdotted,domain=0.505:5]
{sqrt(1 + 1/(4*(x^2-0.5^2)))};
\addlegendentry{intrinsic: \(\sigma_P\neq0\)}

\addplot[thin,dotted,domain=0.25:5]
{1};
\addlegendentry{\(\Delta X_F=\sigma_X\)}

\end{axis}
\end{tikzpicture}
\caption{
Comparison of local, one-sided nonlocal, and two-sided nonlocal uncertainty bounds. The plot is schematic and uses \(\hbar=1\), \(\sigma_X=1\), and \(\sigma_P=0.5\). The local bound falls without limit as momentum width increases. The one-sided nonlocal bound saturates at \(\sigma_X\). The two-sided bound also saturates at \(\sigma_X\) in the ultraviolet but contains the additional threshold \(\Delta P_F>\sigma_P\), reflecting the nonlocal width of momentum.
}
\label{fig:intrinsic-bound-curves}
\end{figure}

In the ordinary local theory the smallest phase-space area is $\hbar/2,$ and in the nonlocal theory this is enlarged to:
\begin{equation}
\frac{\hbar}{2}+\sigma_X\sigma_P.
\end{equation}
The additional term $\sigma_X\sigma_P,$ is the phase-space area contributed by the nonlocal structure of the regulated observable algebra.

This result has an immediate consequences as when both \(\sigma_X\) and \(\sigma_P\) are nonzero, the theory contains a finite phase-space area even beyond the usual quantum contribution. This means that the high energy limit the theory does not admit arbitrarily sharp phase-space points. The obstruction is not merely that a particle cannot be localized at an exact position, but the stronger statement is that a physical excitation cannot be assigned an exact pair $(x,p)$, as a point-supported phase-space datum.

The nonlocal uncertainty principle replaces the local phase-space picture by a nonlocal one, the fundamental object is not a point in phase space, but a finite phase-space cell whose size is controlled by both \(\hbar\) and the nonlocal widths \(\sigma_X\) and \(\sigma_P\). In this way nonlocal quantum field theory implies not only a minimal length scale, but a minimal phase-space resolution.

\section{Squeezed States and the Experimental Question of Resolution}

The existence of some nonlocal phase-space cell seems to suggests to us a natural class of experimental probes, that of which being being squeezed states\footnote{There will be a follow up paper that will fully work the theory behind this experiment. I would like to offer a special thank you to my friend and colleague Hilary Carteret who came up with the idea to use squeezed states as an experiment to test nonlocal quantum field theory.}. The reason is that squeezed states are designed precisely to redistribute uncertainty between conjugate variables so hopefully they can therefore provide us with a direct way to test whether the ordinary local phase-space ellipse can be squeezed indefinitely in one direction, or whether it cannot be squeezed passed a nonlocal width and how the uncertainty relation may scale~\cite{Yuen1976,Walls1983,WallsMilburn1994,
BraunsteinVanLoock2005,Weedbrook2012}.

We will first recall the ordinary squeezed-state structure in one canonical degree of freedom. We first let \(X\) and \(P\) be the ideal local canonical variables satisfying:
\begin{equation}
\notag [X,P]=i\hbar,
\end{equation}
and a minimum-uncertainty Gaussian state satisfies:
\begin{equation}
\notag \Delta X\,\Delta P=\frac{\hbar}{2}.
\end{equation}
A squeezed state is a state for which one of the two variances is reduced while the conjugate variance is increased, in such a way that the product remains at or above the Heisenberg bound. A convenient parametrization we can use is:
\begin{equation}
\Delta X(r)=\Delta X_0 e^{-r},
\qquad
\Delta P(r)=\Delta P_0 e^{r},
\end{equation}
basically a squeeze transformation rescales position and momentum oppositely. In this equation \(r\in\mathbb R\) is the squeezing parameter. The factor of $e^{r}$ is the canonical scaling transformation of the conjugate pair $(X,P)$. The fact that the  the squeeze transformation rescales position and momentum oppositely preserves the commutator because:
\begin{equation}
    [e^{-r}X,e^{r}P]=e^{-r}e^{r}[X,P]=[X,P]=i\hbar,
\end{equation}
so the exponential factors are chosen so that the transformation squeezes one variable while anti-squeezing the conjugate variable by the inverse amount. That is why the uncertainty product remains fixed:
\begin{equation}
\Delta X(r)\Delta P(r)
=
\Delta X_0 e^{-r}\,\Delta P_0 e^{r}
=
\Delta X_0\Delta P_0 .
\end{equation}
Since \(e^{-r}e^{r}=1\), the squeezing of one variable is exactly compensated by the anti-squeezing of its conjugate variable. Here \(\Delta X_0\) and \(\Delta P_0\) are the unsqueezed widths, chosen such that:
\begin{equation}
\Delta X_0\Delta P_0=\frac{\hbar}{2}.
\end{equation}
For \(r>0\), the state is squeezed in position and anti-squeezed in momentum. For \(r<0\), the state is squeezed in momentum and anti-squeezed in position.

In the nonlocal theory the physically relevant widths are not \(\Delta X(r)\) and \(\Delta P(r)\) alone, but rather they are, for position:
\begin{equation}
(\Delta X_F(r))^2
=
(\Delta X(r))^2+\sigma_X^2,
\end{equation}
and for momentum:
\begin{equation}
(\Delta P_F(r))^2
=
(\Delta P(r))^2+\sigma_P^2.
\end{equation}
Substituting the squeezed-state parametrization gives us, for position:
\begin{equation}
(\Delta X_F(r))^2
=
(\Delta X_0)^2 e^{-2r}+\sigma_X^2,
\end{equation}
and now for momentum:
\begin{equation}
(\Delta P_F(r))^2
=
(\Delta P_0)^2 e^{2r}+\sigma_P^2.
\end{equation}
These equations show the characteristic signature of the nonlocal theory, the idea is that in the local theory, increasing \(r\) can make \(\Delta X(r)\) arbitrarily small but in the nonlocal theory:
\begin{equation}
\notag r\to +\infty
\qquad\Longrightarrow\qquad
\Delta X_F(r)\to \sigma_X.
\end{equation}
So arbitrarily strong position squeezing cannot reduce the position width below \(\sigma_X\). Similarly for momentum:
\begin{equation}
\notag r\to -\infty
\qquad\Longrightarrow\qquad
\Delta P_F(r)\to \sigma_P.
\end{equation}
So arbitrarily strong momentum squeezing cannot reduce the momentum width below \(\sigma_P\).

The phase-space product for a squeezed state becomes:
\begin{widetext}
\begin{equation}
\begin{aligned}
\begin{split}
\Delta X_F(r)\Delta P_F(r)
=
\sqrt{(\Delta X_0)^2e^{-2r}+\sigma_X^2}\,
\sqrt{(\Delta P_0)^2e^{2r}+\sigma_P^2}.
\end{split}
\end{aligned}
\end{equation}
\end{widetext}
So before we move on, you may be asking, “But in QFT are these really the variables measured in an experiment?” Our anser is yes.

So we now will show the validity of the squeezed-state test in quantum field theory. We should now explain why the squeezed-state experiment remains a valid test in quantum field theory, and why no additional field-theoretic contact terms have been omitted. The important point is that the experiment does not measure the point fields \(\phi(t,\mathbf x)\) and \(\pi(t,\mathbf y)\) directly. Those are operator-valued distributions satisfying the equal-time canonical commutation relation:
\begin{equation}
\notag[\phi(t,\mathbf x),\pi(t,\mathbf y)]
=
i\hbar\delta^{(3)}(\mathbf x-\mathbf y),
\end{equation}
where the delta function is not an extra physical uncertainty contribution, but it is the distributional contact term which tells us that the fields must be smeared before they become genuine observables.

To start, we let \(f(\mathbf x)\) be the normalized spatial mode selected by the optical, matter-wave, or detector apparatus. We define the corresponding smeared
canonical variables by:
\begin{equation}
Q_f
=
\int d^3x\, f(\mathbf x)\phi(t,\mathbf x),
\end{equation}
and:
\begin{equation}
\Pi_f
=
\int d^3x\, f(\mathbf x)\pi(t,\mathbf x),
\end{equation}
where for simplicity \(f\) is real and normalized according to:
\begin{equation}
\int d^3x\, f^2(\mathbf x)=1.
\end{equation}
Then the field-theoretic delta function is integrated out:
\begin{align}
[Q_f,\Pi_f]
&=
\int d^3x\,d^3y\, f(\mathbf x)f(\mathbf y)
[\phi(t,\mathbf x),\pi(t,\mathbf y)]
\\
&=
i\hbar
\int d^3x\,d^3y\, f(\mathbf x)f(\mathbf y)
\delta^{(3)}(\mathbf x-\mathbf y)
\\
&=
i\hbar
\int d^3x\, f^2(\mathbf x)
\\
&=
i\hbar.
\end{align}
Therefore the selected field mode obeys the ordinary canonical uncertainty relation:
\begin{equation}
\Delta Q_f\,\Delta \Pi_f\geq \frac{\hbar}{2}.
\end{equation}
This is the field-theoretic version of the canonical pair used in the squeezed-state argument. The experiment is therefore not based on a point-particle assumption but it is based on a normalized smeared field mode.

In the nonlocal theory the same mode is acted on by the regulated observable algebra. Fundamentally this means that the measured quadrature distribution is the ideal smeared-mode distribution convolved with the nonlocal response kernel, so:
\begin{equation}
(\Delta Q_{f,F})^2
=
(\Delta Q_f)^2+\sigma_X^2,
\end{equation}
and:
\begin{equation}
(\Delta \Pi_{f,F})^2
=
(\Delta \Pi_f)^2+\sigma_P^2,
\end{equation}
provided the response kernels are normalized, centered, and state-independent. These are exactly the hypotheses used in the variance-addition laws above. The delta-function contact term has therefore already been accounted for by the mode normalization, and it does not generate an additional term in the nonlocal variance law.

It is useful to state the most general possible correction, if the nonlocal kernel were not centered, or if the detector noise were correlated with the state preparation, then the measured variance would take the form:
\begin{equation}
(\Delta Q_{\rm obs})^2
=
(\Delta Q_f)^2+\sigma_X^2+\sigma_{\rm det,X}^2
+2\,{\rm Cov}(Q_f,N_X),
\end{equation}
and similarly:
\begin{equation}
(\Delta \Pi_{\rm obs})^2
=
(\Delta \Pi_f)^2+\sigma_P^2+\sigma_{\rm det,P}^2
+2\,{\rm Cov}(\Pi_f,N_P),
\end{equation}
where \(N_X\) and \(N_P\) denote ordinary experimental noise sources. These terms are not part of the fundamental nonlocal uncertainty relation as they are apparatus-dependent nuisance terms~\cite{BuschLahtiWerner2014}. In a controlled squeezed-state test they must be calibrated, bounded, or subtracted by independent noise characterization. After this calibration, the remaining irreducible floor is the contribution \(\sigma_X^2\) or \(\sigma_P^2\).

Equivalently, in covariance-matrix language, for the canonical quadrature vector~~\cite{MartinMartinez2022,Olivares2012,AdessoRagyLee2014}:
\begin{equation}
Z=(Q_f,\Pi_f)^T,
\end{equation}
a local squeezed state has covariance matrix:
\begin{equation}
\Gamma(r)=S(r)\Gamma_0S(r)^T,
\\ 
S(r)=
\begin{pmatrix}
e^{-r} & 0\\
0 & e^{r}
\end{pmatrix}.
\end{equation}
The nonlocal theory predicts instead:
\begin{equation}
\Gamma_F(r)
=
S(r)\Gamma_0S(r)^T
+
\Sigma_F,
\end{equation}
with:
\begin{equation}
\Sigma_F=
\begin{pmatrix}
\sigma_X^2 & 0\\
0 & \sigma_P^2
\end{pmatrix},
\end{equation}
when the position and momentum response kernels are centered and uncorrelated. If the nonlocal response has a mixed \(XP\) covariance, then we should replace \(\Sigma_F\) by the general matrix:
\begin{equation}
\Sigma_F=
\begin{pmatrix}
\sigma_X^2 & \sigma_{XP}\\
\sigma_{XP} & \sigma_P^2
\end{pmatrix}.
\end{equation}
So again no term is being silently omitted. The diagonal formulas used in this paper correspond to the isotropic or principal-axis case \(\sigma_{XP}=0\). A more general experiment would fit \(\sigma_X\), \(\sigma_P\), and \(\sigma_{XP}\) simultaneously.

The predicted signature is that if the local theory is correct, increasing the squeezing parameter continues to reduce the squeezed quadrature until ordinary technical limits are reached. If the nonlocal phase-space relation is correct, then after calibrated detector noise is removed the squeezed quadrature approaches an irreducible floor:
\begin{equation}
r\rightarrow +\infty
\quad\Longrightarrow\quad
\Delta Q_{f,F}(r)\rightarrow \sigma_X,
\end{equation}
and:
\begin{equation}
r\rightarrow -\infty
\quad\Longrightarrow\quad
\Delta \Pi_{f,F}(r)\rightarrow \sigma_P.
\end{equation}
This is the experimental content of the nonlocal phase-space cell.

So in the local limit this reduces to:
\begin{equation}
\Delta X(r)\Delta P(r)=\frac{\hbar}{2}
\end{equation}
for a minimum-uncertainty squeezed state. With nonlocality present however the product is bounded below by:
\begin{equation}
\Delta X_F(r)\Delta P_F(r)
\geq
\frac{\hbar}{2}+\sigma_X\sigma_P.
\end{equation}
So we find that the nonlocal phase-space cell cannot be squeezed into the ordinary Heisenberg cell. The ellipse may be rotated or stretched, but its fundamental area contains the additional contribution \(\sigma_X\sigma_P\).

This provides a useful experimental interpretation as we may be able to prepare a family of squeezed wave packets with tunable squeezing parameter \(r\), reconstruct their position and momentum widths, and test whether the observed widths follow the local predictions~\cite{ArthursKelly1965}:
\begin{equation}
\notag \Delta X(r)=\Delta X_0e^{-r},
\qquad
\Delta P(r)=\Delta P_0e^{r},
\end{equation}
or instead approach the nonlocal limits:
\begin{equation}
\notag \Delta X_F(r)\to \sigma_X,
\qquad
\Delta P_F(r)\to \sigma_P.
\end{equation}
The clearest signature would be a saturation of the squeezed quadrature width at a nonzero value after known experimental noise sources have been removed.

In the language of quantum optics, \(X\) and \(P\) may be identified with field quadratures~\cite{Husimi1940,Glauber1963,Sudarshan1963,CahillGlauber1969}. In the matter-wave language, they may be identified with position and momentum widths of a spatial wave packet. The present theory is not tied to one platform but what matters is the ability to prepare a family of near-minimum-uncertainty states and vary the squeezing parameter. Squeezed light is already a standard tool in precision interferometry, including gravitational-wave detectors, where it is used to reduce quantum noise below ordinary shot-noise limits~\cite{Caves1981}. This makes squeezed states a natural experimental language for formulating tests of an nonlocal phase-space modification.

The proposed test is conceptually simple, if the local theory is exact, then sufficiently increasing the squeezing parameter should continue to reduce the squeezed quadrature according to the local prediction, limited only by technical noise and ordinary decoherence limits. If the nonlocal phase-space relation is actually correct then after all ordinary noise sources are accounted for the squeezed width should stop decreasing and approach a nonzero floor set by \(\sigma_X\) or \(\sigma_P\), these are functions of the Moffat parameter where $L_M=E_M^{-1}$. This is why we believe that squeezed states will provide us with a direct phase-space probe of the nonlocal cell.

\section{Motion and Wave Structure in the Nonlocal Continuum}

We are now equipped to explain the physical meaning of the nonlocal phase-space cell for mass, velocity, and de Broglie's wavelength. The purpose of this section is to clarify what is changed by the nonlocal theory and what is not changed. In particular the global Poincaré charges may remain well defined, while local densities and point-supported phase-space quantities cease to be fundamental.

We let \(T^{\mu\nu}(x)\) be the stress-energy tensor of the local theory, where \(x=(t,\mathbf x)\in\mathbb R^{1,3}\) is a spacetime point, Greek indices \(\mu,\nu=0,1,2,3\) denote spacetime components, and spatial indices \(i,j=1,2,3\) denote spatial components. The component \(T^{00}(x)\) is the energy density, while \(T^{0i}(x)\) is the \(i\)-th component of the momentum density. In a local relativistic field theory the total four-momentum is defined as:
\begin{equation}
P^\mu
=
\int_{\mathbb R^3}d^3x\,T^{0\mu}(t,\mathbf x),
\end{equation}
where \(P^0\) is the total energy and \(P^i\) are the spatial momentum components. If the theory is translation invariant and the stress-energy tensor is conserved:
\begin{equation}
\partial_\mu T^{\mu\nu}=0,
\end{equation}
then \(P^\mu\) is independent of the time slice.

In the nonlocal theory the local stress-energy tensor is replaced by a regulated stress-energy tensor:
\begin{equation}
T_F^{\mu\nu}(x)
=
F\!\left(\frac{\Box}{E_M^2}\right)T^{\mu\nu}(x),
\end{equation}
where \(F\) is the same entire-function regulator that defines the nonlocal field algebra, \(\Box\) is the d'Alembertian, and \(E_M\) is the Moffat energy scale. Equivalently, we can write:
\begin{equation}
T_F^{\mu\nu}(x)
=
\int d^4y\,K_F(x-y)T^{\mu\nu}(y),
\end{equation}
where \(K_F\) is the spacetime kernel associated with \(F(\Box/E_M^2)\), this equation shows explicitly that the stress-energy assigned to \(x\) is not point-like supported. It is a nonlocal quantity obtained from the stress-energy in a neighbourhood of \(x\) whose characteristic size is set by $L_M\sim1/E_M$.

This distinction is important for the meaning of mass as in particle physics the invariant mass of an asymptotic one-particle state is defined by the Poincaré invariant mass or rest mass of a particle in special relativity:
\begin{equation}
m^2
=
P_\mu P^\mu,
\end{equation}
where \(P^\mu\) is the total four-momentum and the metric convention determines the sign convention. This global invariant mass may remain a good quantum number of an asymptotic state. But mass density or rest-energy density is not a point-supported quantity in the nonlocal theory. The local energy density defined as:
\begin{equation}
\rho(x):=T^{00}(x)
\end{equation}
and it is replaced by:
\begin{equation}
\rho_F(x)
:=
T_F^{00}(x)
=
F\!\left(\frac{\Box}{E_M^2}\right)T^{00}(x).
\end{equation}
So the invariant mass label of an asymptotic particle may remain well defined, while the local mass-energy distribution sourcing fields or curvature is fundamentally nonlocal.

In the gravitational setting this statement becomes especially direct, curvature is sourced by stress-energy so therefore it is trivial to see that if the source is \(T_F^{\mu\nu}\) rather than \(T^{\mu\nu}\), then spacetime curvature is sourced by a nonlocal distribution rather than by an exact point mass. A point-supported source such as:
\begin{equation}
\rho(\mathbf x)=M\delta^{(3)}(\mathbf x)
\end{equation}
is replaced by a smeared density:
\begin{equation}
\rho_F(\mathbf x)
=
F\!\left(\frac{\nabla^2}{E_M^2}\right)\rho(\mathbf x),
\end{equation}
where \(\nabla^2\) is the spatial Laplacian on the equal-time slice. For Gaussian-type regulators this produces a smooth density of width of order the Moffat length \(L_M\). So mass is not fundamentally point-local at the nonlocal scale, even though the total mass:
\begin{equation}
M_F
=
\int d^3x\,\rho_F(\mathbf x)
\end{equation}
may equal the original total mass \(M\), provided the kernel is normalized and suitable boundary conditions hold.

The same logic applies to velocity, when we look at classical mechanics we often write velocity as a point quantity \(v^i=dx^i/dt\), but in a quantum theory velocity is better understood through wave packets, currents, or stress-energy flow. For a relativistic wave packet with central spatial momentum \(\mathbf p\) and mass \(m\), the energy is:
\begin{equation}
E(\mathbf p)
=
\sqrt{|\mathbf p|^2c^2+m^2c^4}.
\end{equation}
In units \(c=1\), this becomes:
\begin{equation}
E(\mathbf p)
=
\sqrt{|\mathbf p|^2+m^2}.
\end{equation}
The corresponding group velocity defined by de Broglie is:
\begin{equation}
\mathbf v_g
=
\nabla_{\mathbf p}E(\mathbf p)
=
\frac{\mathbf p}{E(\mathbf p)}.
\end{equation}
This relation is not discarded since it remains the kinematical relation between the central momentum of a wave packet and the propagation velocity of its phase packet in the infrared.

However we do now see that the velocity field associated with local energy-momentum flow is not point-supported in the nonlocal theory. A natural field-theoretic definition of a local velocity is:
\begin{equation}
v^i(x)
=
\frac{T^{0i}(x)}{T^{00}(x)},
\end{equation}
whenever \(T^{00}(x)\neq 0\). In our nonlocal theory this becomes:
\begin{equation}
v_F^i(x)
=
\frac{T_F^{0i}(x)}{T_F^{00}(x)},
\end{equation}
again assuming \(T_F^{00}(x)\neq 0\). Since both \(T_F^{0i}\) and \(T_F^{00}\) are nonlocal the velocity field \(v_F^i(x)\) is also nonlocal, so velocity is not a sharp property of a point but it is a property of an nonlocal wave packet or energy-momentum flow.

We now turn to the de Broglie's wavelength, for a particle or wave packet with central momentum magnitude:
\begin{equation}
|\mathbf p|=\sqrt{p_1^2+p_2^2+p_3^2},
\end{equation}
the de Broglie wavelength is:
\begin{equation}
\lambda_{\rm dB}
=
\frac{2\pi\hbar}{|\mathbf p|}.
\end{equation}
Equivalently, we can write the wave number as:
\begin{equation}
k
=
\frac{|\mathbf p|}{\hbar},
\end{equation}
we have:
\begin{equation}
\lambda_{\rm dB}
=
\frac{2\pi}{k}.
\end{equation}
The phase of a relativistic plane wave is:
\begin{equation}
\phi(x)
=
-\frac{1}{\hbar}p_\mu x^\mu,
\end{equation}
or, in terms of energy and spatial momentum:
\begin{equation}
\phi(t,\mathbf x)
=
-\frac{Et}{\hbar}
+
\frac{\mathbf p\cdot\mathbf x}{\hbar}.
\end{equation}
Therefore de Broglie's relation remains the relation between momentum and wave phase.

What changes is the interpretation of this wavelength as a resolving scale, we say this as in the local theory by making \(|\mathbf p|\) large makes \(\lambda_{\rm dB}\) small, and we then might expect this to allow arbitrarily fine localization. But in the nonlocal theory though this conclusion fails since the position width is not controlled only by the de Broglie wavelength but it is bounded below by the nonlocal width $\Delta X_F\geq \sigma_X$. Therefore:
\begin{equation}
\lambda_{\rm dB}\to 0
\qquad\not\Rightarrow\qquad
\Delta X_F\to 0,
\end{equation}
a wave may have a very short phase wavelength, but the excitation described by the nonlocal field cannot be assigned an arbitrarily sharp point position.

This gives the new interpretation, de Broglie's wavelength survives as a kinematical phase scale but it is no longer identical to the fundamental resolving power of spacetime. Below the nonlocality scale, the Moffat length \(L_M\), the phase of a wave may continue to oscillate, but the physical excitation remains nonlocal. So the wave is not the wave of an exactly point-localizable particle, it is the wave of an nonlocal field excitation whose position, momentum, mass density, and velocity field are all defined through the regulated observable algebra.

\section{On the Spacetime Meaning of the Nonlocal Description}

We can talk about the spacetime meaning of the nonlocal phase-space relation, so our conclusion is not that spacetime must be discrete, and aslo not that Lorentz covariance must fail. But we say that the conclusion is that exact point-supported spacetime events are not physical elements of the fundamental nonlocal observable algebra.

If we let \(M\) denote the spacetime manifold, in flat spacetime we take:
\begin{equation}
M=\mathbb R^{1,3},
\end{equation}
with coordinates given by:
\begin{equation}
x^\mu=(x^0,x^1,x^2,x^3)=(t,\mathbf x),
\end{equation}
where \(t=x^0\) is the time coordinate and \(\mathbf x=(x^1,x^2,x^3)\) are spatial coordinates. The Minkowski metric is denoted by \(\eta_{\mu\nu}\). With signature \((+,-,-,-)\), the invariant interval between two points \(x,y\in M\) is:
\begin{equation}
(x-y)^2
=
\eta_{\mu\nu}(x-y)^\mu(x-y)^\nu.
\end{equation}
The separation is spacelike when:
\begin{equation}
(x-y)^2<0.
\end{equation}

In the ordinary local field theory we assign observables to exact spacetime points or to arbitrarily small spacetime regions. In a nonlocal field theory this assignment is replaced by the nonlocal map:
\begin{equation}
\notag O(x)
\longrightarrow
O_F(x)
=
F\!\left(\frac{\Box}{E_M^2}\right)O(x),
\end{equation}
or also equivalently:
\begin{equation}
\notag O_F(x)
=
\int d^4y\,K_F(x-y)O(y),
\end{equation}
where again \(O(x)\) is the corresponding local observable of the undeformed theory, \(O_F(x)\) is the regulated nonlocal observable, \(K_F(x-y)\) is the kernel induced by the entire-function regulator, and \(E_M\) is the nonlocal energy scale. The corresponding nonlocal length scale is $L_M\sim E_M^{-1}$, in units where \(\hbar=c=1\).

This equation shows that the object labelled by \(x\) is not supported only at \(x\), it receives contributions from a neighbourhood of \(x\) of characteristic size of the Moffat length \(L_M\)~\cite{PaleyWiener1934,MoffatThompson2026Regulators}. The point \(x\) remains a coordinate on the manifold but it is not the support of an exactly localized physical observable. The physical object is the nonlocal observable \(O_F(x)\), not the ideal point observable of \(O(x)\).

The same conclusion follows from the nonlocal uncertainty relation, we previously found that the nonlocal position and momentum spreads satisfy:
\begin{equation}
\left[(\Delta X_F)^2-\sigma_X^2\right]
\left[(\Delta P_F)^2-\sigma_P^2\right]
\geq
\frac{\hbar^2}{4},
\end{equation}
where \(\Delta X_F\) and \(\Delta P_F\) are the nonlocal position and momentum widths, while \(\sigma_X\) and \(\sigma_P\) are the nonlocal widths of the position and momentum kernels. Since $\Delta X_F\geq \sigma_X$, no physical excitation can be localized below the nonlocal spatial width \(\sigma_X\). So an exact spacetime point cannot be realized as a physical localization state.

This does not require the manifold \(M\) to be replaced by a lattice or any discretization as those have their own issues, but in our case the manifold may remain smooth, and the coordinates \(x^\mu\) may remain useful mathematical labels. What changes is the physical interpretation of those labels, we say that in the local theory one imagines that physical observables can be assigned to arbitrarily small neighbourhoods of a point. In the nonlocal theory, this idealization fails below the Moffat length \(L_M\), so a spacetime point is just a coordinate idealization while a physical event is represented by a nonlocal field excitation.

This distinction also protects Lorentz covariance as since the regulator is built from the covariant operator \(\Box\), the nonlocal smearing can be formulated without choosing a preferred inertial frame~\cite{Thompson2026Covariance}. The theory does not say that spacetime has a preferred lattice spacing in one frame, but actually it says that the algebra of physical observables contains no exact point-supported elements. The nonlocality is intrinsic to the spacetime and fully covariant.

The result may be summarized as follows, the local spacetime picture assumes that one may specify an event by exact data at a point $x^\mu$. The nonlocal theory replaces this by a finite-resolution structure governed by the regulated observable algebra. In phase-space language, we cannot assign exact data $(x,p)$ to a physical excitation. Instead, we have a nonlocal phase-space cell with characteristic widths $\sigma_X,$ and $\sigma_P$. The minimal nonlocal phase-space area is:
\begin{equation}
\notag\Delta X_F\Delta P_F
\geq
\frac{\hbar}{2}
+
\sigma_X\sigma_P.
\end{equation}

So the fundamental conclusion is stronger than a minimal length alone, the nonlocal quantum field theory replaces point spacetime and point phase space by nonlocal spacetime and nonlocal phase space. Exact points remain part of the mathematical continuum, but they are not physical structures in the theory.

\section{Concluding Remarks}

In this paper we have reinterpreted the nonlocal uncertainty relation as the position-sector limit of a broader phase-space statement, the earlier result is not wrong as it can be recovered when position is treated as a nonlocal observable while momentum is kept as the ordinary translation generator. We have extended that reasoning by applying the same nonlocal logic to momentum itself.

The new point we make here is that nonlocal quantum field theory should not be understood just as ordinary quantum theory with imperfect measurements but its physical observables are not exactly point-supported quantities, rather they are nonlocal quantities generated by the regulated field algebra. Once we have accepted this then position and momentum must both be reinterpreted since they are constructed through the finite nonlocal response profiles rather than assigned as exact data at a mathematical point.

The consequence then is that the ordinary Heisenberg principle remains as the local algebraic core of the theory, but it is not the whole physical statement, in the nonlocal theory the physical phase-space cell is enlarged by the nonlocal widths associated with the nonlocal observable algebra. So the ultraviolet theory does not merely prevent arbitrarily sharp localization in space but it prevents the assignment of exact point supported phase-space data altogether.

The lesson of our result is very close in spirit to the original lesson of quantum mechanics but it is sharpened by nonlocal field theory, we find the limitation is not simply a defect in knowledge nor merely a disturbance caused by measurement, but it is a statement about what can count as a physical object in the theory. At sufficiently short distances we find that reality is not described by particles occupying points with definite momenta, but by extended quantum excitations carrying finite phase-space resolution.

The final lesson we should take away from this work is that the fundamental continuum is not necessarily discrete but it is also not physically pointwise. Exact spacetime points and exact phase-space points remain useful mathematical abstractions, but, they are not however physical elements of the fundamental theory.

\section{Acknowledgments}
I would like to thank  my supervisor John Moffat, my friends and colleagues Hilary Carteret and Arvin Kouroshnia, and Robert Mann for helpful discussions on non-local quantum field theory, locality, and measurement in relativistic quantum theory.

\end{document}